\documentclass[5p]{elsarticle}
\usepackage{lineno}
\usepackage[breaklinks=true]{hyperref}
\hypersetup{colorlinks,linkcolor=red,citecolor=blue,urlcolor=blue}

\usepackage{natbib}
\bibpunct{(}{)}{;}{a}{}{,} 

\journal{New Astronomy}

\newcommand{\apjs}{ApJS}
\newcommand{\apj}{ApJ}
\newcommand{\aj}{AJ}
\newcommand{\aaps}{A\&AS}
\newcommand{\aap}{A\&A}
\newcommand{\mnras}{MNRAS}

\newcommand{\pasp}{PASP}

\def\Teff{\ensuremath{T_{\mathrm{eff}}}}

\def\logg{\ensuremath{\log g}}

\def\kms{$\mathrm{km\,s}^{-1}$}
\def\vr{${\upsilon}_{\mathrm{r}}$}
\def\vmic{$\upsilon_{\mathrm{mic}}$}

\def\mbol{$M_{\mathrm{bol}}$}

\begin{document}

\begin{frontmatter}

\title{TYC 3637-1152-1 - a High Amplitude $\delta$ Scuti star with peculiar pulsational properties}


\author[label1]{Ernst Paunzen}
\author[label2,label3]{Klaus Bernhard}
\author[label4]{Moriz Frauenberger}
\author[label4]{Santiago Helbig}
\author[label4]{Andreas Herdin}
\author[label2,label3]{Stefan H{\"u}mmerich}
\author[label1]{Jan Jan{\' i}k}
\author[label4]{Andreas Karnthaler}
\author[label5]{Richard Kom{\v z}{\'i}k}
\author[label4]{Beatrice Kulterer}
\author[label4]{Hans-Michael Maitzen}
\author[label4]{Stefan Meingast}
\author[label4]{Sebastian Miksch}
\author[label5]{Theodor Pribulla}
\author[label4]{Monika Rode-Paunzen}
\author[label4]{Wolfgang Sakuler}
\author[label4]{Carla Schoder}
\author[label6]{Eugene Semenko}
\author[label4]{Nikolaus Sulzenauer}

\address[label1]{Department of Theoretical Physics and Astrophysics, Masaryk University, Kotl\'a\v{r}sk\'a 2, 611 37 Brno, Czech Republic}
\address[label2]{Bundesdeutsche Arbeitsgemeinschaft f{\"u}r Ver{\"a}nderliche Sterne e.V. (BAV), Berlin, Germany}
\address[label3]{American Association of Variable Star Observers (AAVSO), Cambridge, USA}
\address[label4]{Institute for Astrophysics, University of Vienna, T{\"u}rkenschanzstrasse 17, 1180, Vienna, Austria}
\address[label5]{Astronomical Institute, Slovak Academy of Sciences, 059 60 Tatransk{\'a} Lomnica, Slovakia}
\address[label6]{Special Astrophysical Observatory of the Russian Academy of Sciences, Nizhnii Arkhyz 369167, Russia}

\begin{abstract}
In some $\delta$ Scuti stars, only one or two radial modes are excited (usually the fundamental mode and/or first overtone mode) and the observed peak-to-peak amplitudes exceed
0.3\,mag ($V$). These stars are known as High Amplitude Delta Scuti (HADS) variables.

We here present a detailed photometric and spectroscopic analysis of the HADS star TYC 3637-1152-1. We have derived a metallicity close to solar, a spectral type of F4\,V and an age of $\log t = 9.1$. Employing archival time series data from different sources, two frequencies $f_0$\,=\,10.034\,c/d and $f_1$\,=\,12.681\,c/d and their harmonics and linear combinations were identified. The period ratio of $f_0/f_1\,=\,0.791$ puts this star into a peculiar position in the Petersen diagram, from which we conclude that TYC 3637-1152-1 is a unique object with peculiar pulsational properties that indicate a transitional state between HADS stars pulsating in the fundamental and first overtone modes and stars pulsating in higher overtones.
 
\end{abstract}


\end{frontmatter}


\section{Introduction} \label{intro}

The multiperiodic $\delta$ Scuti (DSCT) stars have long been established as a class of variable stars \citep{Fath35}. They usually belong to luminosity classes V to III and are located inside the classical instability strip between spectral types A0 to F5 \citep{Breger98}. $\delta$ Scuti stars show variability on timescales between about 15 minutes and 5 hours \citep{Holds14}, which are caused by multiple radial and non-radial low-order pressure (p) modes that are excited by the $\kappa$ mechanism \citep[e.g.][]{Breger00}.

Several open questions about the nature of $\delta$ Scuti stars remain, e.g. their relationship to the class of $\gamma$ Doradus variables. The theoretical instability regions of both classes on the Hertzsprung-Russell diagram (HRD) overlap \citep{Dupret04}, and hybrid pulsators exhibiting both p and gravity (g) modes are encountered \citep[e.g.][]{Uytt11}. On the basis of space-based data, \citet{Bowman16} discussed the diversity in the pulsational behaviour of $\delta$ Scuti stars, describing several subgroups with 1) constant amplitudes and phases; 2) amplitude modulation caused by the beating of close-frequency pulsation modes; 3) pure amplitude modulation (with no associated phase variation); 4) phase modulation caused by binarity; and 5) amplitude modulation caused by nonlinearity. 

As their name implies, the High-amplitude $\delta$ Scuti (HADS) stars are set apart by the large amplitudes of their light variations ($A_V > 0.3$\,mag) \citep{McNam00}. They usually exhibit one or two stable frequencies \citep{Rodri04} associated with radial pulsation in fundamental or low-order overtone modes \citep{Peter99}. Furthermore, they are usually slow rotators, which seems to be a prerequisite for the observed high-amplitude pulsation \citep{Breger00}. Evidence of period changes in a significant fraction of HADS stars were presented by \citet{Breger98}, who found continuous period decreases and increases, period jumps and even cyclic period variations due to a possible binary light-time effect. However, none of these effects could be correlated with the evolutionary status (i.e. age, life time on the main-sequence, and mass) of the stars \citep{Breger98}.

SX Phoenicis (SX Phe) stars are the Population II counterparts of the $\delta$ Scuti variables \citep{Eggen89} and present phenomenologically similar light curves. Field SX Phe stars generally exhibit the kinematic properties of halo (or thick disk) stars, asymmetric and large-amplitude light curves, and low metallicities \citep{McNam95}. From studies of globular clusters, it was found that SX Phe stars have pulsation amplitudes $A_V > 0.1$\,mag, and simple light curves consistent with radial pulsation and one or two dominant frequencies. However, as more accurate photometric data became available, SX Phe variables with much lower amplitudes have also been discovered \citep{Cohen12}. Current pulsation models imply that the classical $\kappa$ mechanism in the He\,II partial ionization zone is the corresponding driving mechanism in these stars \citep{Fiore14}.

For both HADS and SX Phe stars, period-luminosity (PL) relations have been established \citep{Peter99,Cohen12} which makes them interesting targets for distance estimations. Here, we present a detailed photometric and spectroscopic study of TYC\,3637-1152-1, a neglected HADS variable, investigating its location in the HRD and its pulsation characteristics.

\begin{table}[t]
\caption{Observed quantities and astrophysical parameters of TYC\,3637-1152-1. Kinematic data and radial velocity information were taken from Gaia DR2.}
\label{parameters}
\begin{center}
\begin{tabular}{lc}
\hline
Parameter & Value \\
\hline
RA (J2000) & 23h\,30m\,37s.26 \\ 
Dec (J2000) & +46$^\circ$\,24'\,04''.3 \\
$l$ (deg) & 108.7086 \\
$b$ (deg) & $-$14.2266 \\
$D$ (pc) & 525(15) \\
$Z$ (pc) & $-$129(15) \\
$V$ (mag) & 10.398(47) \\
$E(B-V)$ (mag) & 0.10(2) \\
\Teff\ (K) & 6\,800(250) \\
\mbol\ (mag) & 1.46(8) \\
$\pi$ (mas) & 1.905(54) \\
$\mu_\alpha \cos \delta$ (mas\,yr$^{-1}$) & $-$6.998(74) \\
$\mu_\delta$ (mas\,yr$^{-1}$) & +15.697(68) \\
\vr\ (\kms) & $-$21.73(4.76) \\
$U$ (\kms) & $-$16 \\
$V$ (\kms) & +5 \\
$W$ (\kms) & +53 \\
$T$ (\kms) & 56 \\
\hline   
\end{tabular}    
\end{center}                                      
\end{table}

\begin{figure}[t]
 \begin{center}
		\includegraphics[width=0.45\textwidth]{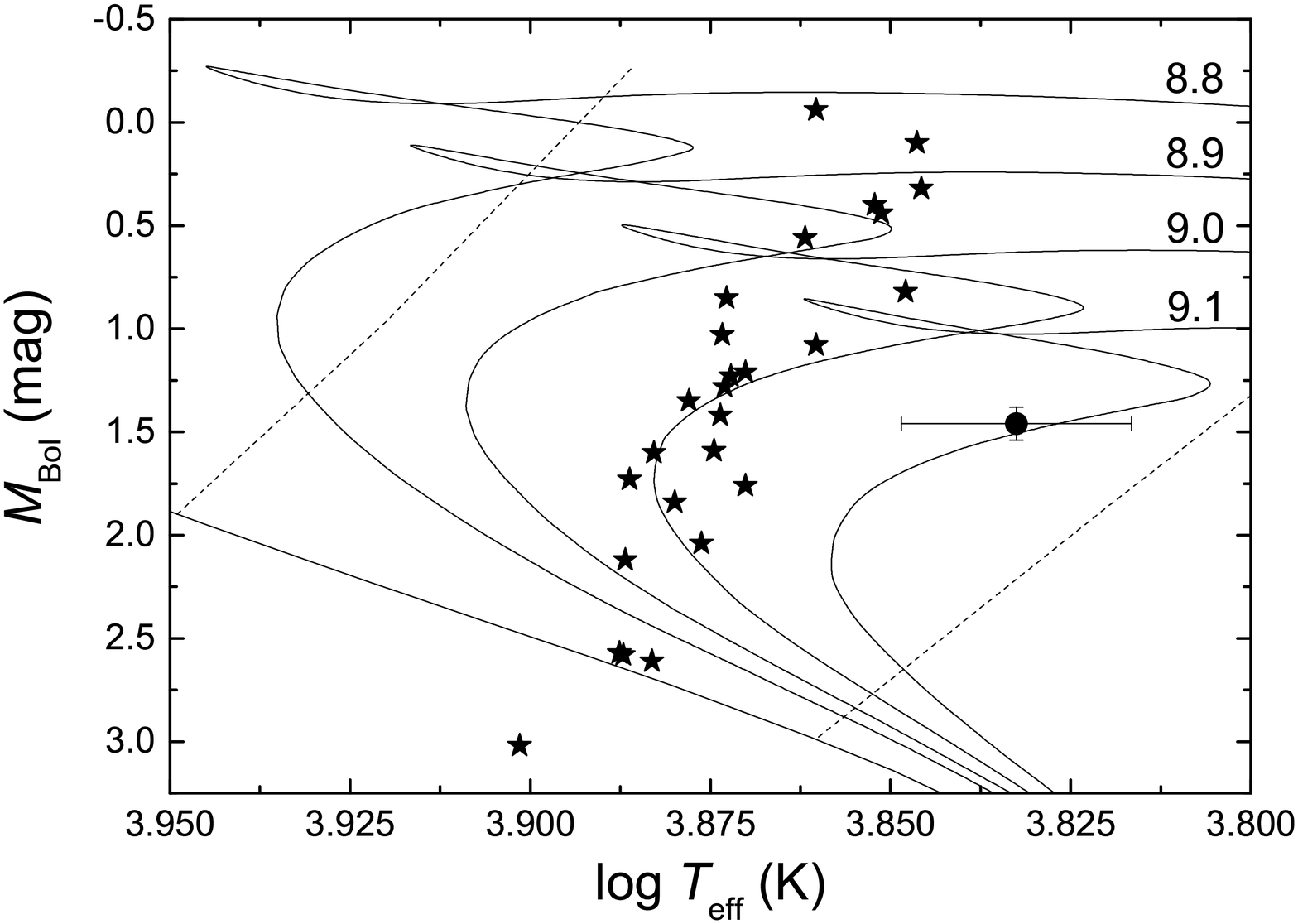}
   \caption[]{The location of TYC\,3637-1152-1 in the $M_{\rm Bol}$ versus $\log T_\mathrm{eff}$ diagram. The asterisks denote the positions of HADS variables from \citet{McNam97}. The dashed lines indicate the borders of the $\delta$\,Scuti instability strip according to \citet{Breger98}. Also shown are isochrones for solar metallicity ([Z]\,=\,0.019) and different ages, which have been taken from the PARSEC database \citep{Bressan12}.}
   \label{hrd}
 \end{center}
\end{figure}

\section{Target star and its astrophysical parameters}

The variability of TYC\,3637-1152-1 (NSVS 3564994, V0670 And) was first reported by \citet{Dimit07}, who used Northern Sky Variability Survey (NSVS) data to search for variable stars in the Andromeda constellation. The star was classified as a $\delta$ Scuti variable with a period of 0.09966\,d and an amplitude of 0.29\,mag.\footnote{NSVS data consist of unfiltered CCD observations that have been placed onto a $V$-equivalent scale; the resulting passband is approximately 4\,000--9\,000\,\AA\ \citep{Aker00} (cf. Section \ref{survey_data}).} Since then, no further studies of TYC\,3637-1152-1 have been published.

As an initial step, we determined our target star's kinematic properties and location in the HRD to decide whether it is a Population I (HADS) or II (SX Phe) star. The observed quantities and the final results are listed in Table \ref{parameters}.

Data Release 2 (DR2) of the $Gaia$ satellite mission\footnote{http://vizier.u-strasbg.fr/viz-bin/VizieR-3?-source=I/345/gaia2} lists a parallax of $\pi$\,=\,1.905(54)\,mas for our target star. This converts to a distance between 510 and 540\,pc, respectively. Employing the 3D dust maps of the Pan-STARRS1 project \citep{Green18}, we find that TYC\,3637-1152-1 is located in a region only marginally affected by interstellar absorption and deduce a reddening $E(B-V)$ between 0.09 and 0.11\,mag for the above listed distance range. In the following, we have therefore adopted a value of 0.31\,mag for the total absorption. Because neither observations in Johnson $U$ nor in Str{\"o}mgren $u$ \citep{Paunzen15} are available, we have not been able to apply any additional photometric dereddening procedure. 

Using parallax, total absorption and bolometric correction from \citet{Flower96} and an apparent magnitude of $V = 10.398(47)$\,mag \citep{Henden16}, we have derived a bolometric magnitude of \mbol\,=\,1.46(8)\,mag for TYC\,3637-1152-1. For the estimation of \Teff, we employed photometry from APASS DR9 \citep{Henden16} and the 2MASS catalogue \citep{Skru06} and scaled the total absorption to the corresponding filter bands. As next step, independent photometric calibrations for different colours \citep{Pinso12,Peca13,Paunzen17} were employed. All these calibrations yield a consistent value of 6\,800(250)\,K. To test this \Teff, the VOSA (VO Sed Analyzer) tool v5.1 \citep{Bayo08} was applied to fit the Spectral Energy Distribution (SED) to the available photometry. The best fit was achieved for a main-sequence 6\,750\,K SED of solar metallicity, supporting our results.

Figure \ref{hrd} illustrates the location of TYC\,3637-1152-1 and other HADS variables \citep{McNam00} in the $M_{\rm Bol}$ versus $\log T_\mathrm{eff}$ diagram. Also indicated are isochrones for different ages and solar metallicity ([Z]\,=\,0.019) taken from the PARSEC database \citep{Bressan12}. With an age of $\log t = 9.1$, our target lies well within the borders of the $\delta$\,Scuti instability strip \citep{Breger98}, although it is significantly cooler than the other HADS variables -- a fact which might also have a bearing on its pulsational behaviour (cf. Sect. \ref{analyse}).

Galactic $UVW$ space velocities and the total space velocity $T$ (often referred as ``Toomre'' energy) are important and useful for discriminating halo, thick disk and thin disk stars \citep{Sandage87}. According to \citet{Fuhrm04}, thin disk stars have $T$ values up to 50\,\kms, whereas a mixture between thin and thick disk stars is observed between 50 and 75\,\kms. A value of 180\,\kms\ separates thick disk from halo stars.

The list of SX Phe variables published by \citet{Nemec17}, for example, includes one star with $T = 144$\,\kms, while all other stars have much higher values up to 1\,274\,\kms. We here employ a left-handed Galactic system, i.e. $U$ positive in the anti-centre direction, $V$ positive in the direction of Galactic rotation, and $W$ positive in the direction of the North Galactic Pole. Calculations are done according to the formulae presented by \citet{John87} with respect to the Local Standard of Rest \citep{Cosku11}.

The space velocity values for TYC\,3637-1152-1 are listed in Table\,\ref{parameters}. Radial velocity information and proper motions were taken from Gaia DR2. The derived $T$ value of 56\,\kms\ is not in agreement with a Population II object and strongly supports that our target star belongs to Population I and hence is an HADS star. Proper motion and distance $Z$ to the Galactic plane (Table \ref{parameters}) are compatible with the kinematics of the thin disk within the solar neighbourhood \citep{Pasetto12}. The $W$ component (+53\,\kms) is rather large compared to other stars in the solar vicinity \citep{Holmb09} but still within the 5$\sigma$ range of the velocity distribution for the thin disk \citep{Anguiano17}. 

\begin{figure}[t]
 \begin{center}
		\includegraphics[width=0.5\textwidth]{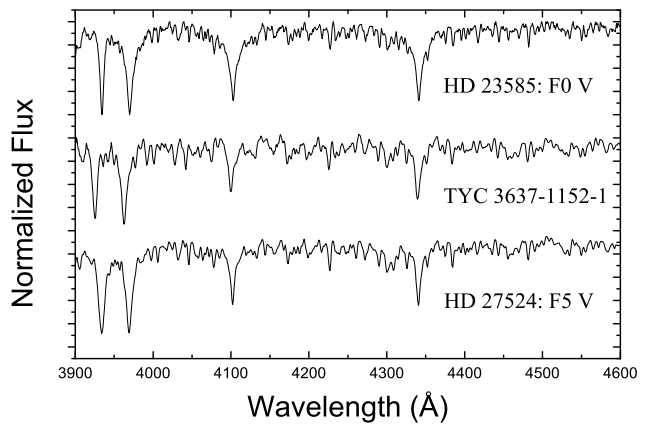}
   \caption[]{The SAO classification resolution spectrum of TYC\,3637-1152-1 in comparison to two MKK standard stars.}
   \label{spectra}
 \end{center}
\end{figure}

To sum up, from its kinematic properties, we conclude that TYC\,3637-1152-1 belongs to Population I and is an HADS variable.

\section{Spectroscopic observations} \label{spectr}

To shed more light on the nature of TYC\,3637-1152-1, spectroscopic observations have been secured. Both high and classification resolution spectroscopy guarantee the best possible analysis. Details on the spectroscopic observations and employed instrumentations are provided below.

\begin{itemize}
\item 0.8\,m Cassegrain telescope at Vienna University Observatory (Austria), LISA spectrograph (Shelyak), covering 3\,900 to 8\,000\,\AA\ with a resolving power of about 1\,500, S/N of about 20, spectrum taken during night of 19./20.01.2018
\item 1.0\,m telescope of the Special Astrophysical Observatory (SAO, Russia), UAGS spectrograph, covering 3\,900 to 5\,300\,\AA\ with a resolving power of about 5\,000, S/N of about 100, spectrum taken during night of 05./06.06.2018
\item 1.3\,m telescope at Tatransk{\'a} Lomnica (Slovakia), MUSICOS spectrograph, covering 4\,200 to 7\,300\AA\ 
with a resolving power of about 35\,000, S/N of about 25, spectrum taken during night of 19./20.02.2018
\end{itemize}

The reduction of the raw frames and extraction of the 1D spectra using the IRAF package tasks, Linux shell scripts, and FORTRAN programs has been described by \citet{Pribulla15}.

The classification resolution spectrum of TYC\,3637-1152-1 was classified using MKK standards as proposed by \citet{Gray89}. The S/N ratio of the spectrum is sufficient to classify the spectral type (temperature) and the overall appearance of the metallic line spectrum, i.e. a Population II object can be identified with certainty. We classify the spectrum as F4\,V (Fig. \ref{spectra}) which is in line with the star's location in the HRD (Fig. \ref{hrd}). The observed metallic line spectrum is as strong as in the standard stars.

To investigate the metallicity of TYC\,3637-1152-1 in more detail, we computed a synthetic spectrum using the program SPECTRUM\footnote{http://www.appstate.edu/$\sim$grayro/spectrum/spectrum.html} \citep{Gray94} and modified versions of the ATLAS9 code taken from the Vienna New Model Grid of Stellar Atmospheres, NEMO\footnote{http://www.univie.ac.at/nemo} \citep{Heiter02}. We used a stellar atmosphere with the following parameters: \Teff\,=\,6800\,K, \logg\,=\,3.8, and \vmic\,=\,2\,km\,s$^{-1}$. The synthetic spectrum was first folded with the instrumental profile and then with different rotational profiles, which yielded a best fit for 50\,km\,s$^{-1}$, with an uncertainty of about 5\,km\,s$^{-1}$. To test these parameters, a grid of atmospheres with effective temperatures and surface gravities around the input values were applied. The H$\alpha$ and H$\beta$ lines are best fitted with the original values (\Teff, \logg, and \vmic\ as listed above) under the constraint that they are not sensitive to \logg. To estimate the [M/H] value, we used different models from +0 to $-$2\,dex, as compared to solar values. We have investigated the region between 5\,000 and 5\,600\,\AA\ because it boasts the highest S/N ratio and the most prominent metallic lines. Within the noise level, the solar abundance synthetic spectrum reproduces the observed spectrum very well. From a comparison of synthetic spectra with $-$0.2 and $-$0.5\,dex, we estimate
that the overall elemental abundance of TYC\,3637-1152-1 is not lower than $-$0.2\,dex.

In summary, the star's spectroscopic characteristics corroborate that it belongs to Population I.

\begin{figure}[t]
 \begin{center}
		\includegraphics[width=0.5\textwidth]{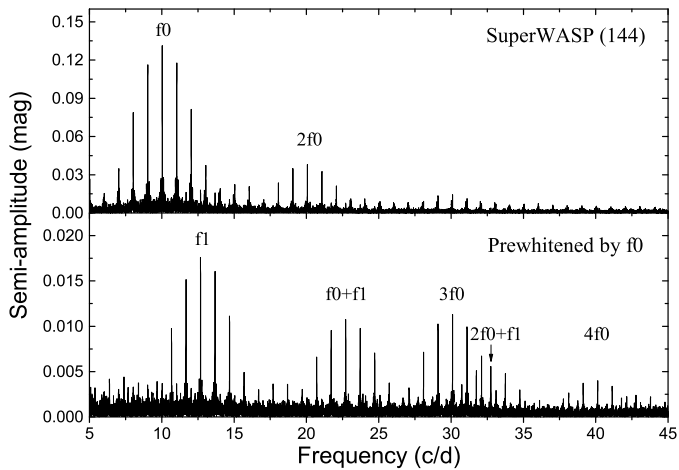}
   \caption[]{Frequency analysis of one SuperWASP data set (CCD camera \#144) for TYC\,3637-1152-1 in the investigated range of $5 < f(c/d) < 50$. The upper and lower panels illustrate the Fourier spectra for unwhitened data and data that has been prewhitened with $f_0$, respectively.}
   \label{amp_spec}
 \end{center}
\end{figure}

\begin{figure}[t]
 \begin{center}
		\includegraphics[width=0.5\textwidth]{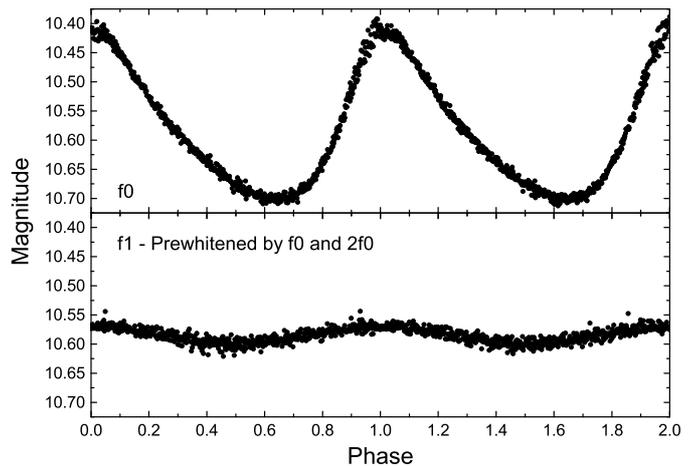}
   \caption[]{The phased light curves of TYC\,3637-1152-1, based on the SuperWASP data set for CCD camera \#143 (Table \ref{stars_tsa}).}
   \label{phased_light_curve}
 \end{center}
\end{figure}

\begin{figure}[t]
 \begin{center}
		\includegraphics[width=0.5\textwidth]{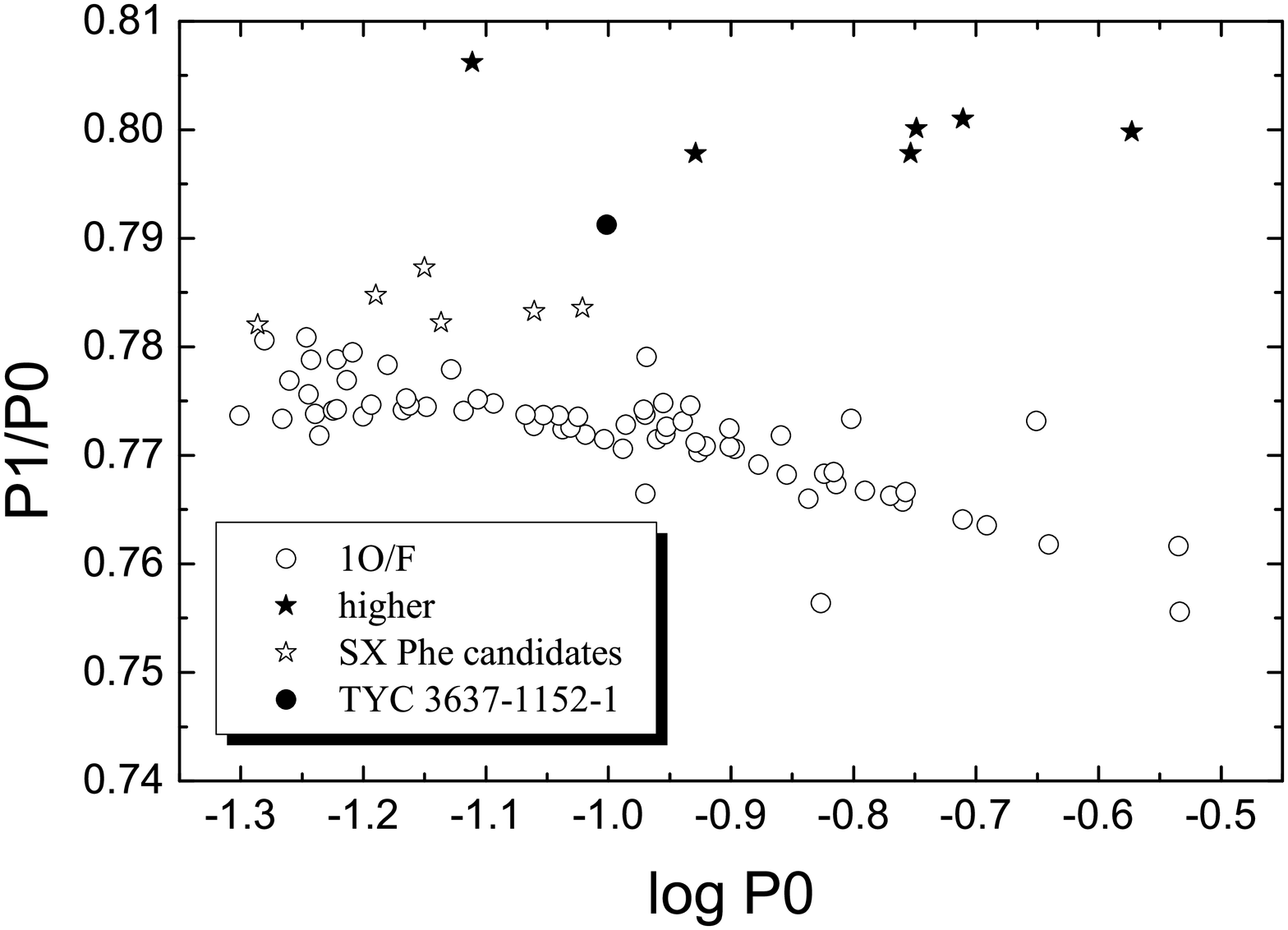}
   \caption[]{Petersen diagram for double-mode HADS stars after \citet{Furgoni16}. Open circles denote stars pulsating in the fundamental and first overtone modes. Filled asterisks refer to stars pulsating in higher overtone modes. These are BD+08 4583 \citep{Khrus11}, HD 77906 \citep{Khrus11}, HD 146953 \citep{Khrus09}, V798 Cyg \citep{Musaz98}, V1719 Cyg \citep{Poret88}, and VZ Cnc \citep{Fu99}. Open asterisks are possible SX Phe star candidates (Section \ref{analyse}). The position of TYC 3637-1152-1 is indicated by the filled circle. For the sake of clarity, the star TYC 2706-1244-1 \citep{Khrus14} with $P1/P0$\,=\,0.834 (probably pulsating in the second and third overtones) has not been included.}
   \label{petersen}
 \end{center}
\end{figure}

\section{Photometric data sources and time series analysis} \label{survey_data}

In order to investigate the light variability of our target star, photometric observations from the All-Sky Automated Survey for Supernovae (ASAS-SN), Northern Sky Variability Survey (NSVS), and Wide Angle Search for Planets (SuperWASP) archives were procured.

The ASAS-SN survey is imaging the entire visible sky every night to a depth of $V$\,$<$\,17\,mag \citep{Kocha17}. The available data span up to five years of observations. As of end-2017, ASAS-SN consists of five stations equipped with four 14\,cm aperture Nikon telephoto lenses. Observations are made using $V$ (two stations) or $g$ (three stations) band filters and three dithered 90\,s exposures. ASAS-SN saturates at 10 to 11\,mag; the exact limit depends on the camera and the image position. However, a procedure inherited from the ASAS survey is applied which corrects for saturation but introduces additional noise in the corresponding data sets \citep{Pojma02}.

NSVS data were collected with the ROTSE-I instruments \citep{Wozn04} and comprise unfiltered CCD observations that have been placed onto a $V$-equivalent scale \citep{Aker00}; the resulting passband is approximately $4\,000-9\,000$\,\AA. The time base is about one year and 100 to 500 measurements are typically available per object.

The SuperWASP survey started in 2004 and covers both hemispheres. It provides long-term photometric time series in a broadband filter (4\,000 -- 7\,000\,\AA) with an accuracy better than 1\% for objects in the magnitude range 8\,$<$\,$V$\,$<$\,11.5\,mag \citep{Polla06}. Observations consist in general of two consecutive 30\,s integrations followed by a 10-minute gap. Here, we have used data from the first and only WASP public data release \citep{Butte10}\footnote{https://wasp.cerit-sc.cz/}. Measurements with an error larger than 0.05\,mag have been excluded, and each camera has been treated separately.

All data sets with more than 100 points were cleaned of obvious outliers by visual inspection and searched for periodic signals in the frequency range of $5 < f(c/d) < 50$ using \textsc{Peranso} \citep{Paunz16} and \textsc{Period}04 \citep{Lenz05}. Consecutive prewhitening with the most significant frequency was applied until no significant frequencies remained. Neither time binning nor smoothing was applied to the data sets.

\begin{table*}[t]
\caption{Characteristics of the employed time series (time baseline and number of observations) and the derived frequencies and amplitudes along with their error estimates.}
\label{stars_tsa}
\begin{center}
\begin{tabular}{lccccc}
\hline
Source & Frequency & Semi-Amplitude & Time-base & $N$ & Designation \\
& (c/d) & (mag) & (d) & & \\
\hline
ASAS-SN	&	10.034019(1)	&	0.1543(5)	&	1653.27057	&	671	&	$f_0$	\\
	&	20.068035(4)	&	0.0391(5)	&		&		&	2$f_0$	\\
	&	12.681113(9)	&	0.0182(5)	&		&		&	$f_1$	\\
	&	30.10203(1)	&	0.0118(5)	&		&		&	3$f_0$	\\
	&	22.71513(1)	&	0.0113(5)	&		&		&	$f_0$+$f_1$	\\
	&	32.74916(2) &	0.0071(5)	&		&		&	2$f_0$+$f_1$	\\ 
	& 39.13593(3) & 0.0045(5) &   &   & 4$f_0$-1\\
NSVS (set1)	&	10.03378(9)	&	0.099(4)	&	260.73834	&	286	&	$f_0$	\\
	&	20.0676(2)	&	0.038(4)	&		&		&	2$f_0$	\\
NSVS (set2)	&	10.0354(3)	&	0.108(12)	&	226.76294	&	228	&	$f_0$	\\
SWASP (143)	&	10.03403(1)	&	0.1325(3)	&	139.90112	&	5424	&	$f_0$	\\
	&	20.06804(3)	&	0.0369(3)	&		&		&	2$f_0$	\\
	&	12.68123(7)	&	0.0173(3)	&		&		&	$f_1$	\\
	&	30.1019(1)	&	0.0111(3)	&		&		&	3$f_0$	\\
	&	22.7152(1)	&	0.0102(3)	&		&		&	$f_0$+$f_1$	\\
	& 32.7498(3) & 0.0045(3) &   &   & 2$f_0$+$f_1$ \\
	& 40.1356(3) & 0.0042(3) &   &   & 4$f_0$ \\
SWASP (144)	&	10.03404(1)	&	0.1310(4)	&	133.90130	&	5064	&	$f_0$	\\
	&	20.06786(4)	&	0.0364(4)	&		&		&	2$f_0$	\\
	&	12.68105(9)	&	0.0177(4)	&		&		&	$f_1$	\\
	&	30.1020(1)	&	0.0116(4)	&		&		&	3$f_0$	\\
	&	22.7155(1)	&	0.0108(4)	&		&		&	$f_0$+$f_1$	\\  
	& 32.7492(3) & 0.0051(4) &   &   & 2$f_0$+$f_1$ \\
	& 40.1354(4) & 0.0040(4) &   &   & 4$f_0$ \\
\hline   
\end{tabular}    
\end{center}                                      
\end{table*}

\begin{table*}[t]
\caption{Astrometric and kinematic data, and $G$ magnitudes from Gaia DR2 for HADS stars with $P1/P0$ ratios larger than 0.782. The ranges for the distance from the Sun ($D$) and the Galactic plane ($Z$) were calculated taking into account the error of the parallax ($\pi$).}
\label{new_sxphes}
\begin{center}
\footnotesize
\begin{tabular}{lcccccccc}
\hline
Star & $G$ & $\pi$ & $\mu_\alpha \cos \delta$ & $\mu_\delta$ & $l$ & $b$ & Range -- $D$ & Range -- $Z$ \\
& (mag) & (mas) & (mas\,yr$^{-1}$) & (mas\,yr$^{-1}$) & (deg) & (deg) & (kpc) & (kpc) \\
\hline
GSC 00010--00276 & 13.894 & 0.5011(359) & +3.745(77) & $-$32.766(43) & 116.3717 & $-$62.0453 & 1.9 -- 2.2 & 1.3 -- 1.5 \\
GSC 07460--01520 & 14.461 & 0.2467(433) & +2.015(60) & +3.504(37) & 10.1198 & $-$34.7526 & 3.5 -- 4.9 & 0.7 -- 1.0 \\
LINEAR 2653935 & 16.397 & 0.2703(961) & $-$6.941(163) & $-$3.481(97) & 270.2379 & +65.6724 & 2.7 -- 5.7 & 0.8 -- 1.7 \\
LINEAR 9328902 & 16.136 & 0.1041(477) & $-$14.761(91) & $-$9.442(43) & 34.2640 & +80.1150 & 6.6 -- 17.7 & $-$6.6 -- $-$17.7 \\
LINEAR 16586778 & 15.794 & 0.1859(410) & $-$2.552(62) & $-$9.655(73) & 46.9936 & +45.5337 & 4.4 - 6.9 & 4.4 -- 6.9 \\
NSV 7805 & 16.709 & 0.1822(856) & +3.096(185) & $-$4.637(99) & 13.2194 & +29.3222 & 3.7 -- 10.4 & $-$3.2 -- $-$9.0 \\
\hline   
\end{tabular}    
\end{center}                                      
\end{table*}

\section{Analysis and Conclusions} \label{analyse}

Table \ref{stars_tsa} lists the results of the time series analysis of the employed data sets. The detection of the frequencies $f_0$ (10.034\,c/d) and $f_1$ (12.681\,c/d) is completely without doubt. Furthermore, harmonics and linear combinations of these frequencies are also clearly present. {Interestingly, no harmonics of $f_1$ in any of the data sets was detected.
Fig. \ref{amp_spec} exemplarily illustrates the Fourier spectrum of the SuperWASP data set for CCD camera \#144 in the investigated frequency range of $5 < f(c/d) < 50$. In Fig. \ref{phased_light_curve}, the corresponding phased light curves are shown.

For the two main frequencies, we calculate a ratio of $f_0/f_1\,=\,0.791$. This is well above the theoretical limit (0.775) for stars pulsating in the fundamental and first overtone modes 
but also below the value (0.780) for the first and second overtone \citep{Peter96}. To further investigate this matter, we have employed the models by \citet{Stell79} and calculated the mean density ($\log \rho/\rho_{\odot} = -1.2$) of TYC\,3637-1152-1 using the stellar parameters listed in Table \ref{parameters}. For the given mean density, the observed frequencies almost 
exactly match the first and second overtones from the models.

We also have to emphasize that non-radial modes were found in HADS and SX Phe stars \citep{Zhou99,Poret11}. Normally, they have smaller amplitudes than radial modes and have been
only observed together with radial modes. We cannot rule out that $f_1$ is a non-radial mode, but it is highly unlikely. For a definite identification, multicolour photometry or time resolved
high resolution spectroscopy is needed. 

Seven HADS stars are known that do not pulsate in the fundamental and first overtone modes. The observed period ratios for these objects range from 0.797 to 0.834. Only three of these stars have been studied in more detail, viz. V798 Cyg \citep{Musaz98}, V1719 Cyg \citep{Poret88}, and VZ Cnc \citep{Fu99}. Only basic data like periods and amplitudes are available for the other four stars, viz. BD+08 4583 \citep{Khrus11}, HD 77906 \citep{Khrus11}, HD 146953 \citep{Khrus09}, and TYC 2706-1244-1 \citep{Khrus14}. 

As an aside, we would like to comment on the possible misclassification of some stars currently included in the group of HADS variables. The following five stars show light variations with amplitudes less than 0.3\,mag in $V$: BD+08 4583 (semi-amplitude of the main period, 0.071\,mag), HD 77906 (0.128\,mag), HD 146953 (0.072\,mag), TYC 2706-1244-1 (0.093\,mag), and V798 Cyg (0.091\,mag). Therefore, according to the most commonly used definition (cf. Sect. \ref{intro}), these stars do not qualify as HADS variables. In this respect, it is interesting to point out that \citet{Musaz98} write in their introduction that HADS stars are defined by amplitudes larger than $\sim$0.2\,mag ($V$). As reference, they cite \citet{Peter99}, who, however, do not comment on a lower amplitude limit but state that HADS stars exhibit amplitudes of 0.4\,mag ($V$). Even if we adopt a limit of $\sim$0.2\,mag, out of the above-listed objects, only HD 77906 would qualify as an HADS star. This demonstrates the need for a critical assessment of all published
HADS variables, which might eventually lead to a refinement of the definition of HADS and SX Phe stars. This, however, is out of the scope of the present study.

Figure \ref{petersen} shows the Petersen diagram for double-mode HADS stars taken from \citet{Furgoni16}, who compiled a catalogue of stars classified as such in the AAVSO International Variable Star Index (VSX) \citep{Watson06} by the end of December 2015. We did not check the amplitudes of the stars from this catalogue and whether they qualify as 
HADS variables by the standard criteria. The \citet{Furgoni16} sample was expanded by the stars apparently pulsating in higher overtones discussed above. The only exception is TYC 2706-1244-1, which has been excluded for the sake of clarity. With a period ratio of $P1/P0$\,=\,0.834, this star might be pulsating in the second and third overtones, as indeed suggested by \citet{Khrus14}.

The general trend of the pulsational behaviour as described in \citet{Poret05} -- an almost linear behaviour of $P1/P0$ versus $\log P0$ -- is clearly visible. Interestingly, our target star lies in the transition region between objects pulsating in the fundamental and first overtone modes and objects pulsating in higher overtones, which makes it quite unusual.

The models predict an upper limit for $P1/P0$ of about 0.775 for $\log P0 < -0.85$ and solar metallicity. Stars of lower metallicity, e.g. Population II objects, tend to have larger $P1/P0$ values up to 0.780. Six objects boast $P1/P0$ values larger than 0.782 and are therefore promising SX Phe star candidates: GSC 00010-00276, GSC 07460-01520, LINEAR 2653935, LINEAR 9328902, LINEAR 16586778, and NSV 7805. For LINEAR 9328902 and LINEAR 2653935, light curves are available in the VSX which clearly demonstrate an HADS star-like variability.

To investigate this matter, we have procured parallax information and proper motions from Gaia DR2. The astrometric and kinematic results are summarized in Table \ref{new_sxphes} and indicate that all these stars belong to Population II and are consequently SX Phe variables. The observed high $P1/P0$ values are therefore due to the lower metallicity of these objects and are not caused by high overtone mode pulsation.

This scenario, however, is not suited to explain the peculiar position in the Petersen diagram of the HADS star TYC\,3637-1152-1, which is of near solar abundance (cf. Sect. \ref{spectr}). Employing the available PL relations, we have calculated absolute or bolometric magnitudes. We note that the bolometric correction from \citet{Flower96} accounts for only $\sim$0.02\,mag, and thus does not have a big impact on the derived values. First of all, we used the relations for SX Phe stars derived by \citet{Santo01}. Assuming that $f_0$ is the fundamental mode and using the observed effective temperature, we derive $M_{\rm Bol} = 1.6$\,mag, which is in the range of the here derived value of 1.46(8)\,mag (cf. Table \ref{parameters}).

Assuming that $f_0$ is the first overtone mode, we are faced with the problem that, to our knowledge, no PL relation for the first overtone periods of $\delta$ Scuti or HADS stars exists. We have therefore employed the PL relation of \citet{McNam00} for the fundamental pulsation mode and scaled $f_0$ with 0.78. From this, we get $M_{\rm V} = 1.4$\,mag, which agrees well with the here derived value of 1.46(8)\,mag. If we assume that $f_0$ is the fundamental mode, we derive $M_{\rm V} = 1.8$\,mag is not
in line with the observation. This is another strong argument that $f_0$ is indeed the first overtone mode.
We thus conclude that TYC\,3637-1152-1 follows the classical PL relation and it is not outstanding in that respect.

In summary, we conclude that with a metallicity close to solar, a spectral type of F4\,V and an age of $\log t = 9.1$, TYC 3637-1152-1 is an unique object with peculiar pulsational properties that indicate a transitional state between HADS stars pulsating in the fundamental and first overtone modes and stars pulsating in higher overtones. This result is supported by the star's position in the HRD and Petersen diagram.

From an assessment of the literature, it is also evident that the definition of the HADS and SX Phe variables has to be verified on the basis of newly available photometric, astrometric and kinematic data, and there is the need for a critical assessment of all published HADS variables.

\section*{Acknowledgements}
\small

This project was supported by the grants 7AMB17AT030 (M\v{S}MT).
This paper makes use of data from the DR1 of the WASP data \citep{Butte10} as provided by the WASP consortium, 
and the computing and storage facilities at the CERIT Scientific Cloud, reg. no. CZ.1.05/3.2.00/08.0144 
which is operated by Masaryk University, Czech Republic.
This article was created by the realisation of the project ITMS No.
26220120029, based on the supporting operational Research and development
program financed from the European Regional Development Fund.

\end{document}